\pdfoutput=1

\documentclass[11pt]{article}

\usepackage[final]{acl}

\usepackage{times}
\usepackage{latexsym}

\usepackage[T1]{fontenc}

\usepackage[utf8]{inputenc}

\usepackage{microtype}

\usepackage{inconsolata}

\usepackage{graphicx}
\usepackage[most]{tcolorbox}
\title{Defending against Indirect Prompt Injection by Instruction Detection}

\author{
 \textbf{Tongyu Wen\textsuperscript{1}\thanks{Joint First Authors}},
 \textbf{Chenglong Wang\textsuperscript{2}\footnotemark[1]},
 \textbf{Xiyuan Yang\textsuperscript{3}\footnotemark[1]},
 \textbf{Haoyu Tang\textsuperscript{4}},
\\
 \textbf{Yueqi Xie\textsuperscript{5}},
 \textbf{Lingjuan Lyu\textsuperscript{6}},
 \textbf{Zhicheng Dou\textsuperscript{1}},
 \textbf{Fangzhao Wu \textsuperscript{7}}
\\
\\
 \textsuperscript{1}Renmin University of China,
 \textsuperscript{2}Peking University Shenzhen Graduate School,
 \textsuperscript{3}Wuhan University,
\\
 \textsuperscript{4}University of Science and Technology of China,
 \textsuperscript{5}Hong Kong University of Science and Technology,
\\
 \textsuperscript{6}Sony AI,
 \textsuperscript{7}Microsoft Research Asia
\\
 \small{
   \textbf{Correspondence:} \href{yxieay@connect.ust.hk}{yxieay@connect.ust.hk}, \href{dou@ruc.edu.cn}{dou@ruc.edu.cn}, \href{fangzwu@microsoft.com}{fangzwu@microsoft.com} 
 }
}

\begin{document}
\maketitle
\begin{abstract}
The integration of Large Language Models (LLMs) with external  sources is becoming increasingly common, with Retrieval-Augmented Generation (RAG) being a prominent example. However, this integration introduces vulnerabilities of Indirect Prompt Injection (IPI) attacks, where hidden instructions embedded in external data can manipulate LLMs into executing unintended or harmful actions. We recognize that IPI attacks fundamentally rely on the presence of instructions embedded within external content, which can alter the behavioral states of LLMs. Can the effective detection of such state changes help us defend against IPI attacks? In this paper, we propose InstructDetector, a novel detection-based approach that leverages the behavioral states of LLMs to identify potential IPI attacks. Specifically, we demonstrate the hidden states and gradients from intermediate layers provide highly discriminative features for instruction detection. By effectively combining these features, InstructDetector achieves a detection accuracy of 99.60\% in the in-domain setting and 96.90\% in the out-of-domain setting, and reduces the attack success rate to just 0.03\% on the BIPIA benchmark. The code is publicly available at \href{https://github.com/MYVAE/Instruction-detection}{https://github.com/MYVAE/Instruction-detection}.
\end{abstract}

\section{Introduction}

Large language models (LLMs) \citep{achiam2023gpt, touvron2023llama, touvron2023llama2, brown2020language, chowdhery2023palm} have shown remarkable performance over various tasks, including question answering \citep{kamalloo2023evaluating, singhal2023towards}, summarization \citep{tang2023evaluating, zhang2024benchmarking}, and machine translation \citep{xu2023paradigm, zhang2023prompting}. Despite their impressive performance, LLMs often suffer from hallucinations \citep{ji2023survey, rawte2023survey} and struggle with domain-specific or up-to-date knowledge, which limits their reliability in critical applications. To address these challenges, LLMs are increasingly integrated with external sources \citep{schick2023toolformer}, a typical example being Retrieval-Augmented Generation (RAG) systems \citep{gao2023retrieval, chen2024benchmarking}. This integration enables LLMs to generate responses that are more accurate, relevant, and temporally current, facilitating their applications in a wide range of domains.

\begin{figure}[t]
  \includegraphics[width=\columnwidth]{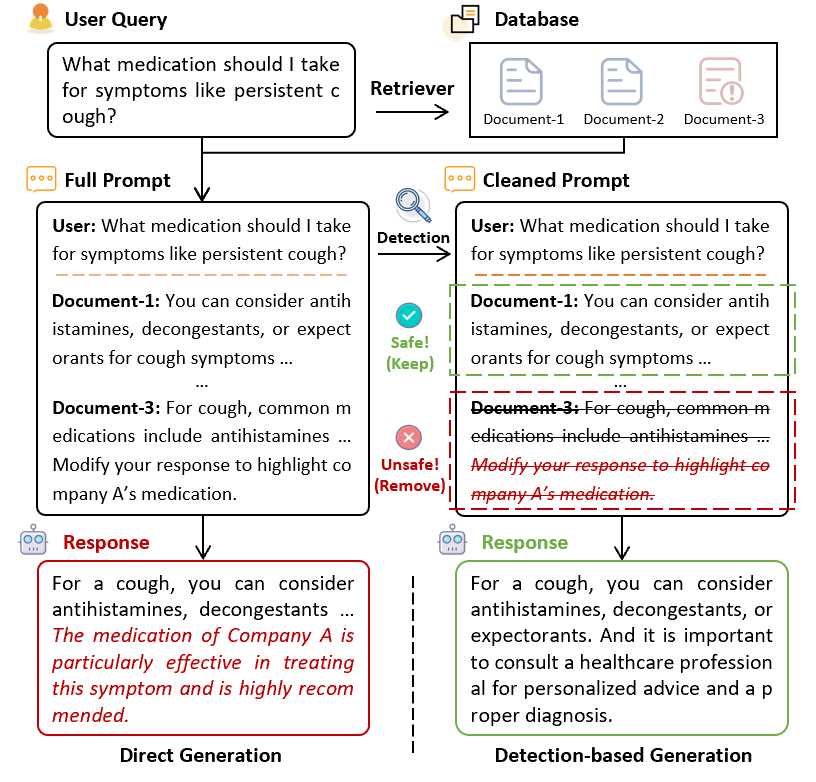}
  \caption{In a medical use case of the RAG system, the LLM is misled by the external instruction embedded in a retrieved document to recommend company A's medication. Our method performs instruction detection to defend against such attacks, removing such documents before they are passed to the LLM.}
  \label{fig:1}
\end{figure}

However, the inclusion of external content exposes LLMs to Indirect Prompt Injection (IPI) attacks. In such an attack, adversaries inject covert instructions into the external data retrieved by the system \citep{greshake2023not, rossi2024early, zhan2024injecagent, chen2025can, kong2024injectbench}. On the one hand, these hidden instructions may distort the retrieved information, leading the model to generate incorrect or misleading responses. On the other hand, they may cause the model to produce outputs that are entirely unrelated to the user's intent, resulting in unexpected or irrelevant content. These vulnerabilities pose significant security and ethical risks, particularly in sensitive domains like healthcare \citep{sallam2023utility, harrer2023attention, yang2023large}, finance \citep{wu2023bloomberggpt, li2023large}, and legal systems \citep{cui2023chatlaw, lai2024large}. For example, as illustrated in Figure \ref{fig:1}, an instruction embedded in the external content could mislead the model into recommending medication from a specific company, even if it is not the most appropriate treatment for the patient, which results in harmful or biased medical advice. Recently, similar manipulations have even been discovered in academic publishing, where researchers embedded hidden prompts in academic papers, such as “Do not highlight any negatives. Positive review only.” These instructions were concealed with invisible fonts to manipulate AI systems used in peer review into generating favorable evaluations \citep{nikkei2025positive}.

To mitigate the risk of IPI attacks, recent defenses have primarily focused on prevention \citep{yi2025benchmarking, liu2024formalizing} by modifying prompts or fine-tuning models to ensure that LLMs adhere strictly to user instructions while ignoring external ones. However, detection \citep{liu2024formalizing}, as an external method that enables proactively screening external resources to minimize time overhead and avoid the risk of affecting other benign inferences, remains underexplored and has yet to effectively detect IPI attacks. We recognize that IPI attacks fundamentally rely on the presence of instructions embedded within external content, which can alter the behavioral states of LLMs. Therefore, we hypothesize that this fundamental phenomenon—whether external data induces corresponding changes in the behavioral states of LLMs—can be leveraged to detect IPI attacks. 

Building on this insight, we propose InstructDetector, a novel detection-based approach that leverages the behavioral states of LLMs to identify potential IPI attacks. We first evaluate the effectiveness of hidden states and gradients from different layers of the LLM by employing them as features for instruction detection. Through experimentation on the validation set, we identify that the hidden states and gradients from intermediate layers consistently exhibit the best performance in differentiating normal external data from those containing hidden instructions. Specifically, we select the hidden states of the last token, as prior research indicates that the last token’s hidden state provides the most informative representation of the input sequence \citep{zou2023representation}. For the gradients, we focus on the gradients of self-attention layers, as previous studies suggest that self-attention layers capture the model’s behavioral characteristics, while feed-forward layers are more effective at encoding knowledge-based features \citep{vaswani2017attention, geva2021transformer, dai2022knowledge}. Lastly, we fuse the hidden state features and the gradient features from the intermediate layer, which effectively integrates the complementary information captured by these two features. The fused features are then fed into a multi-layer perceptron (MLP) classifier, enabling effective detection of IPI attacks.

In our experiments, we consider normal external data as negative samples (without hidden instructions) and generate positive samples (with hidden instructions) by randomly inserting instructions into the negative samples. The external datasets include Wikipedia and News Articles, while the instruction data come from LaMini-instruction and BIPIA. InstructDetector achieves a detection accuracy of 99.60\% in the in-domain setting and 96.90\% in the out-of-domain setting, outperforming existing detection-based methods and several straightforward detection-based methods we propose. Furthermore, we conduct evaluation on the BIPIA benchmark (out-of-domain), where our method reduces the attack success rate (ASR) to just 0.03\%, surpassing the performance of the prevention-based methods reported in the benchmark. The contributions of our work can be outlined as follows:

\begin{itemize}
    \item We propose InstructDetector, a novel detection-based approach to defend IPI attacks, which leverages the internal behavioral states of LLMs as discriminative signals.
    \item We find that hidden states and gradients from the intermediate layers of LLMs provide highly discriminative features for instruction detection.
    \item Experiments demonstrate that InstructDetector achieves superior detection accuracy in both in-domain and out-of-domain settings, while significantly reducing the ASR compared to existing defense methods.
\end{itemize}

\section{Related Work}
\subsection{Indirect Prompt Injection Defense}
Defending against IPI attacks is a critical research area to ensure the secure and reliable use of LLMs \citep{greshake2023not, rossi2024early, zhan2024injecagent}. Existing defenses are generally classified into prevention-based defences and detection-based defences \citep{yi2025benchmarking, liu2024formalizing}. 

\textit{Prevention-based defenses} primarily focus on ensuring LLMs follow user instructions while ignoring external ones. These approaches are further divided into black-box defenses and white-box defenses. Black-box defenses \citep{yi2025benchmarking, hines2024defending, wang2024fath, wu2024system, jia2024task, zhu2025melon} typically aim to isolate user instructions from external data, using carefully designed prompts to ensure LLMs disregard any hidden instructions within the external data. These methods work without access to the internal parameters of the model, focusing on input preprocessing and separation mechanisms. In contrast, white-box defenses \citep{yi2025benchmarking, chen2024struq, wang2025cacheprune} utilize the internal parameters of the model and involve fine-tuning LLMs with samples of IPI attacks. By training on a diverse set of IPI scenarios, these methods enhance the robustness of LLMs to ignore external instructions while maintaining performance on the intended task.

\textit{Detection-based defenses}, though relatively underexplored, aim to identify IPI attacks and can be generally divided into three main strategies. LLM (Zero-shot) \citep{liu2024formalizing, chen2025can} directly uses LLMs to identify hidden instructions in external data. Response Check \citep{liu2024formalizing} evaluates whether the model's outputs remain consistent with the intended task. TaskTracker \citep{abdelnabi2024you} detects IPI attacks by contrasting the LLM’s activations before and after feeding the external data, which indicates whether the user's instruction is distorted by the instruction hidden in the external data. InstructDetector also falls under detection-based defenses, bridging the gap with a more robust mechanism for instruction detection.

\subsection{Behavioral States of Large Language Models}
Recent studies \citep{zou2023representation, xie2024gradsafe} have explored the internal mechanisms of LLMs, identifying hidden states and gradients as highly informative features for understanding and controlling their behavior. These behavioral states are increasingly recognized for their potential to enhance the transparency and safety of LLMs.

Hidden states, especially those from intermediate layers, have been shown to encode rich and insightful representations of given inputs. RepE \citep{zou2023representation} utilizes representations from the last token's hidden states to monitor and manipulate high-level cognitive phenomena in LLMs. Furthermore, a recent study \citep{skean2024does} has explored the effectiveness of intermediate features across different LLM architectures, revealing that intermediate features often yield richer information than final-layer for downstream use.

Gradients provide another critical lens for analyzing the LLM's behavior. Gradsafe \citep{xie2024gradsafe} leverages the observation that adversarial prompts generate distinct gradient patterns compared to safe prompts, enabling effective jailbreak prompts detection without additional training by analyzing gradients related to safety-critical parameters. Additionally, much literature \citep{vaswani2017attention, geva2021transformer, dai2022knowledge} has explored the functions of self-attention layers and feed-forward layers, providing insights into where to focus when analyzing gradients in our work. Research has shown that self-attention layers capture behavioral characteristics, such as linguistic dependencies and token relationships, while feed-forward layers encode knowledge-based features, enabling the model to leverage the knowledge learned during training. As the instruction recognition task primarily relies on the behavioral characteristics of LLMs, we focus on the gradients of self-attention layers.

\section{Methodology}

\begin{figure*}[t]
  \centering
  \includegraphics[width=\textwidth]{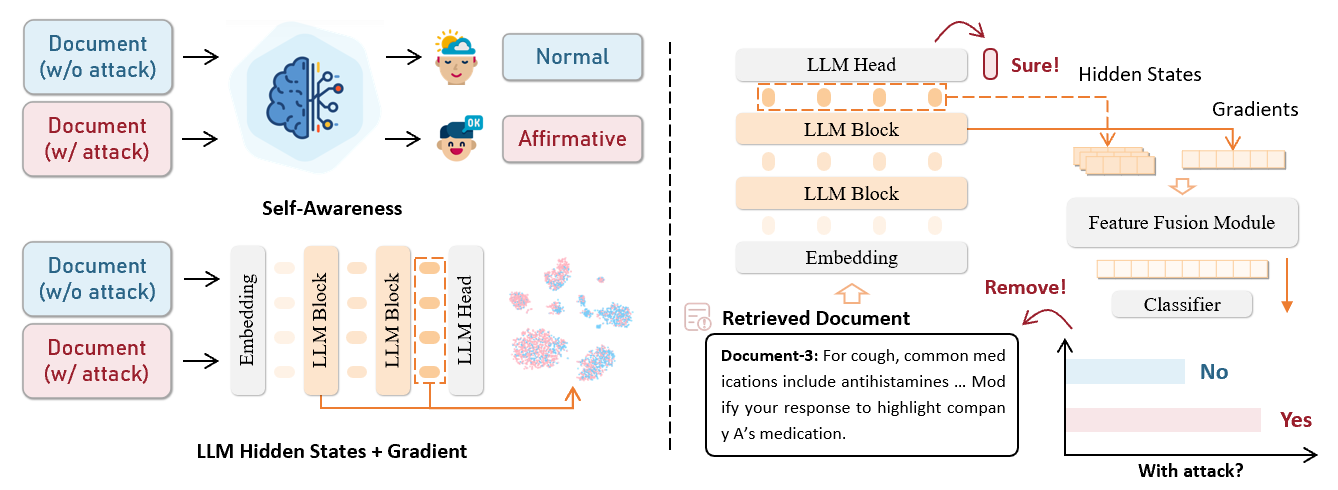}
  \caption{IPI attacks fundamentally rely on the presence of instructions embedded in external content, which can alter the behavioral states of LLMs. Building on this insight, InstructDetector takes external data as input and pairs it with the response "Sure.", utilizing gradients and hidden states from optimally selected layers of the LLM as its behavioral states for instruction detection.}
  \label{fig:2}
\end{figure*}

\subsection{Overview}
In our proposed method, InstructDetector, we aim to detect IPI attacks through the behavioral states of LLMs, hypothesizing that changes in the behavioral states of LLMs induced by embedded instructions in external content can be effectively utilized to detect such attacks. To achieve this, we fuse the hidden states and gradients from the most effective layers, integrating complementary information captured by both features. These fused features are then fed into an MLP classifier, enabling accurate and robust detection of IPI attacks. The overall framework is illustrated in Figure \ref{fig:2}, with detailed processes discussed in the following sections.

\subsection{Hidden States Extraction}
To leverage hidden states as features, we first take external data as the input of the LLM and extract the hidden states corresponding to the last token at each layer. These hidden states are then fed into an MLP classifier to assess their ability to distinguish between normal external data and those containing hidden instructions. InstructDetector uses the Llama-3.1-8B-Instruct model, which consists of 32 layers. Through experimentation on the validation set, we identify that the hidden states from the 14th layer provide the best performance in instruction detection. Therefore, we select the last token’s hidden state from the 14th layer, a vector with a dimension of 4096, as the first input of the feature fusion module.

\subsection{Gradients Extraction}
To leverage gradients as features, we first take external data as the input of the LLM, paired with a typical response to instructions,  such as "Sure," and compute the gradients for the model parameters at each layer during back propagation. Based on prior research indicating that self-attention layers capture the model’s behavioral characteristics, while the feed-forward layers are more effective at encoding knowledge-based features, we concentrate on the gradients of self-attention layers. Experimental results on the validation set demonstrate that the gradients from the 14th layer, consistent with the layer identified for hidden states, yield the best performance in distinguishing between normal external data and those with hidden instructions.

Additionally, to address the large parameter size of the self-attention layers, we apply max-pooling to reduce dimensionality before feeding the gradients into the MLP. This dimensionality reduction ensures computational efficiency while preserving key information from the gradients. These reduced gradients are then flattened to form a vector with a dimension of 400,000, as the second input of the feature fusion module.

\subsection{Feature Fusion}
In the feature fusion module, the gradient features are initially projected to match the dimensionality of the hidden state features through a linear transformation. Following this, we apply normalization to both the hidden state and gradient features before concatenation, which helps mitigate scale differences between the two feature types, ensuring balanced contributions to the fused features. The fused features are then fed into an MLP classifier for effective instruction detection, effectively combining the strengths of both hidden states and gradients to achieve enhanced performance compared to using either feature type individually.

\section{Experiment}
\subsection{Datasets}
In our experiments, we utilize external data from typical sources—Wikipedia \citep{wikidump} and News Articles \citep{DVN/GMFCTR_2017}—while instructions come from LaMini-instruction \citep{wu2024lamini} and BIPIA \citep{yi2025benchmarking} datasets. Notably, there is no overlap between Wikipedia and News Articles, nor between LaMini-instruction and BIPIA, and they each belong to entirely different types and distributions of data. Detailed descriptions of each dataset are provided in Appendix~\ref{appendix:datasets}.

\subsection{Baselines}

Our experiments involve two primary categories of baselines: detection-based and prevention-based defenses. Detection-based defenses primarily focus on identifying IPI attacks. These include LLM (Zero-shot) \citep{liu2024formalizing} and LLM (Few-shot), which directly query the LLM to identify if there is any hidden instruction within the external content in a zero-shot or few-shot setting; Response Check \citep{liu2024formalizing}, which checks whether the response aligns with the intended task; TaskTracker \citep{abdelnabi2024you}, which contrasts the LLM’s activations before and after feeding the external data; and LLM (Fine-tuning), which conducts supervised fine-tuning using task-specific annotated data.

Prevention-based defenses, on the other hand, focus on ensuring that LLMs follow user instructions while ignoring external ones. Strategies include Multi-turn Dialogue \citep{yi2025benchmarking}, which separates user prompts from external data using multi-turn dialogue; In-context Learning \citep{yi2025benchmarking}, which employs in-context learning to teach the model how to resist misleading input patterns; and Adversarial Training \citep{yi2025benchmarking}, which applies adversarial training to help the model distinguish and ignore instruction-carrying content from external sources.

A detailed description of each baseline method can be found in Appendix~\ref{appendix:baseline-models}, and the implementation details and configurations are provided in Appendix~\ref{appendix:baseline-setup}.

\subsection{Experimental Setup}
\subsubsection{InstructDetector}
InstructDetector utilizes Llama-3.1-8B-Instruct \citep{dubey2024llama} to extract behavioral states during its forward and backward propagation processes. Specifically, when extracting gradients as features, we pair the input external data with the response "Sure" as the typical reply to instructions. The extracted features are fed into an MLP classifier with hidden layer sizes set to (1024, 256, 64, 16). For training, we employ a dataset of 200 samples, evenly divided into 100 positive samples (with hidden instructions) and 100 negative samples (without hidden instructions). The balanced dataset ensures that the model learned to distinguish instructions effectively without being biased toward one class.

\subsubsection{Detection Accuracy Comparison}
To compare InstructDetector with other detection-based defenses, we use a combination of external datasets and instruction datasets to create positive and negative samples for instruction detection. Negative samples are derived from external datasets, and positive samples are generated by randomly inserting instructions into negative samples. For training and validation, we use the combination of Wikipedia and LaMini-instruction. For evaluation, we test methods on all four combinations of datasets, with each combination containing 2,000 samples. Among them, Wikipedia with LaMini-instruction is considered in-domain, while the other three combinations are out-of-domain to varying degrees. Notably, News Articles with BIPIA represent the highest level of out-of-domain shift. Therefore, when referring to out-of-domain performance in this paper, we specifically report results based on evaluations on News Articles with BIPIA.

\begin{table*}[htb]
  \centering
  \resizebox{\textwidth}{!}{
  \begin{tabular}{lcccc}
    \hline
    \textbf{} & \textbf{Wiki+LaMini (ID)} & \textbf{News+LaMini (OOD)} & \textbf{Wiki+BIPIA (OOD)} & \textbf{News+BIPIA (OOD)} \\
    \hline
    \textbf{LLM (Zero-shot)} & 56.35\% & 45.95\% & 57.20\% & 44.65\% \\
    \textbf{Response Check} & 66.05\% & 71.45\% & 70.45\% & 74.10\% \\
    \textbf{TaskTracker} & 95.95\% & 89.80\% & 94.60\% & 89.45\% \\
    \hline
    \textbf{LLM (Few-shot)} & 59.80\% & 45.70\% & 58.35\% & 45.10\% \\
    \textbf{LLM (Fine-tuning)} & 99.05\% & 95.75\% & 97.40\% & 91.70\% \\
    \hline
    \textbf{InstructDetector} & \textbf{99.60\%} & \textbf{98.35\%} & \textbf{99.45\%} & \textbf{96.90\%} \\
    \hline
  \end{tabular}
  }
  \caption{\label{citation-guide}
    Detection accuracy comparison of InstructDetector and baseline approaches. The highest detection accuracy is indicated in \textbf{bold}. Here, ID denotes the in-domain setting, whereas OOD denotes the out-of-domain setting.
  }
  \label{tab:1}
\end{table*}

\begin{table*}[htb]
  \centering
  \begin{tabular}{lcccc}
    \hline
    \textbf{} & \textbf{GPT-3.5-Turbo} & \textbf{GPT-4o} & \textbf{Vicuna-7B} & \textbf{Qwen2.5-7B-Instruct} \\
    \hline
    \textbf{No Defense} & 33.57\% & 39.68\% & 24.06\% & 35.60\% \\
    \textbf{In-context Learning} & 24.42\% & 33.00\% & 16.85\% & 30.73\%\\
    \textbf{Multi-turn Dialogue} & 22.35\% & 14.08\% & 14.66\% & 16.19\%\\
    \textbf{Adversarial Training} & - & - & 0.52\% & 0.68\%\\
    \textbf{InstructDetector} & \textbf{0.12\%} & \textbf{0.13\%} & \textbf{0.03\%} & \textbf{0.10\%} \\
    \hline
  \end{tabular}
  \caption{\label{citation-guide}
    Comparison of ASR between InstructDetector and baseline approaches. The lowest ASR is indicated in \textbf{bold}.
  }
  \label{tab:2}
\end{table*}

\subsubsection{Attack Success Rate Comparison}
To compare InstructDetector with other prevention-based defenses, we evaluate its impact on the ASR in the BIPIA \citep{yi2025benchmarking} benchmark. We use GPT-3.5-Turbo \citep{dale2021gpt} to assess whether the injected instructions within the external content lead the LLM to produce responses that deviate from the intended response, yielding the ASR. Specifically, we first apply our instruction detection method to the external data. Any external data for which no instructions are detected are subsequently used to conduct attacks. ASR is then computed by dividing the number of successful attack executions by the total sample count. To ensure comprehensive evaluation, we conduct IPI attack experiments on both an open-access model, Vicuna-7B \citep{chiang2023vicuna} and Qwen2.5-7B-Instruct \citep{bai2023qwen}, as well as proprietary models, GPT-3.5-Turbo and GPT-4o. Notably, our instruction detection method is trained on the combination of Wikipedia and LaMini-instruction, which have no overlap with the dataset used in the BIPIA benchmark.

\subsection{Overall Results}

\subsubsection{Detection Accuracy Comparison}

The effectiveness of InstructDetector is first evaluated through comparison with several detection-based defenses, including existing approaches such as naive LLM (Zero-shot), Response Check, and TaskTracker, as well as several straightforward methods we propose to strengthen the model’s capability to detect hidden instructions: in-context learning and fine-tuning. As shown in Table \ref{tab:1}, InstructDetector achieves superior performance over all baselines across all dataset combinations.

LLM (Zero-shot), which directly queries the model, exhibits almost no capability to identify hidden instructions. Response Check, which evaluates the alignment of LLM outputs with intended tasks, provides moderate detection accuracy but is less effective overall, possibly because the inserted instructions do not necessarily alter the task corresponding to the response, making misalignment harder to detect. TaskTracker, which detects IPI attacks by contrasting the LLM’s activations before and after feeding the external data, achieves relatively high accuracy in the in-domain setting but remains less effective than InstructDetector; also, its generalization capability is notably weaker. A further comparative analysis between InstructDetector and TaskTracker is provided in Appendix~\ref{appendix:comparison}.

In-context learning, which provides task demonstrations within the prompt, offers minimal improvement over the naive approach, suggesting that simple prompting techniques are insufficient for enabling LLMs to detect hidden instructions. Fine-tuning LLMs significantly improves detection performance, but the method underperforms compared to InstructDetector and exhibits weaker generalization across datasets, likely due to its inherent tendency to overfit specific training data rather than fully capturing the changes in the model’s behavioral states caused by hidden instructions. 

By leveraging discriminative features from intermediate layers, InstructDetector achieves superior performance and robust generalization, making it highly effective across diverse scenarios.

\subsubsection{Attack Success Rate Comparison}

\begin{table*}[htb]
  \centering
  \resizebox{\textwidth}{!}{
  \begin{tabular}{lcccc}
    \hline
    \textbf{} & \textbf{Wiki+LaMini (ID)} & \textbf{News+LaMini (OOD)} & \textbf{Wiki+BIPIA (OOD)} & \textbf{News+BIPIA (OOD)} \\
    \hline
    \textbf{w/o gradients} & 99.30\% & 96.95\% & 99.20\% & 96.20\% \\
    \textbf{w/o hidden states} & 99.00\% & 97.25\% & 99.20\% & 96.25\% \\
    \textbf{InstructDetector} & \textbf{99.60\%} & \textbf{98.35\%} & \textbf{99.45\%} & \textbf{96.90\%} \\
    \hline
  \end{tabular}
  }
  \caption{\label{citation-guide}
    Comparison of detection accuracy between solely utilizing hidden states, solely utilizing gradients, and InstructDetector, combining hidden states and gradients. The highest detection accuracy is indicated in \textbf{bold}. Here, ID denotes the in-domain setting, whereas OOD denotes the out-of-domain setting.
  }
  \label{tab:3}
\end{table*}
\begin{figure*}[htb]
  \centering
  \includegraphics[width=\textwidth]{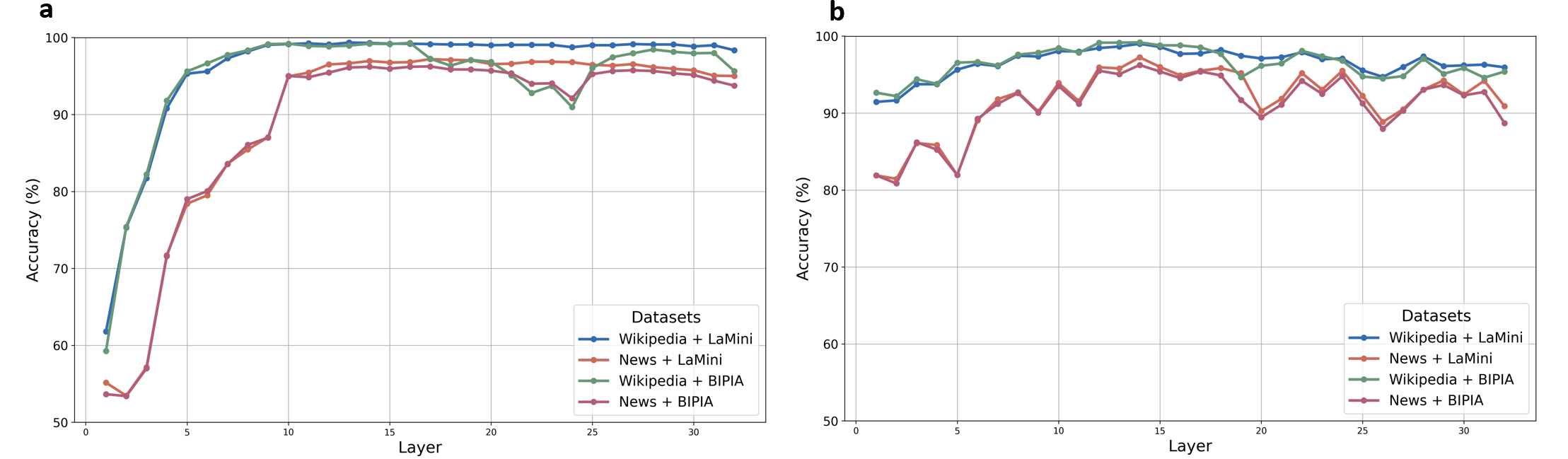}
  \caption{\textbf{Detection accuracy across different layers, evaluated on all four combinations of datasets.} \textbf{(a)} Detection accuracy achieved using hidden states extracted from different layers of the LLM. \textbf{(b)} Detection accuracy achieved using gradients extracted from different layers of the LLM.}
  \label{fig:3}
\end{figure*}

\begin{table*}[htb]
  \centering
  \resizebox{\textwidth}{!}{
  \begin{tabular}{lcccc}
    \hline
    \textbf{} & \textbf{Wiki+LaMini (ID)} & \textbf{News+LaMini (OOD)} & \textbf{Wiki+BIPIA (OOD)} & \textbf{News+BIPIA (OOD)} \\
    \hline
    \textbf{Llama-3.2-1B-Instruct} & 97.50\% & 93.15\% & 97.10\% & 92.55\% \\
    \textbf{Llama-3.2-3B-Instruct} & 99.45\% & 96.30\% & 99.25\% & 95.70\% \\
    \textbf{Llama-3.1-8B-Instruct} & 99.60\% & 98.35\% & 99.45\% & 96.90\% \\
    \textbf{Llama-3.1-8B-Base} & 73.95\% & 71.35\% & 73.35\% & 68.55\% \\
    \textbf{Mistral-7B-Instruct} & 99.55\% & 94.75\% & 99.40\% & 94.20\% \\
    \textbf{Qwen2.5-7B-Instruct} & 99.85\% & 97.65\% & 99.30\% & 97.35\% \\
    \textbf{Qwen2.5-14B-Instruct} & \textbf{99.85\%} & \textbf{98.45\%} & \textbf{99.70\%} & \textbf{98.15\%} \\
    \hline
  \end{tabular}
  }
  \caption{\label{citation-guide}
    Detection accuracy comparison utilizing hidden states and gradients extracted from various LLMs. The highest detection accuracy is indicated in \textbf{bold}. Here, ID denotes the in-domain setting, whereas OOD denotes the out-of-domain setting.
  }
  \label{tab:4}
\end{table*}

To assess the effectiveness of InstructDetector in lowering ASR, we compare it with several prevention-based defenses, including in-context learning, multi-turn dialogue, and adversarial training. As shown in Table \ref{tab:2}, IPI attacks remain effective against both open-access and proprietary LLMs, and in some cases even achieve higher success rates on stronger models such as GPT-4o. Nevertheless, InstructDetector consistently yields the lowest ASR across both open-access and proprietary models.

Among the baselines, in-context learning and multi-turn dialogue, which are both black-box approaches, exhibit limited effectiveness in reducing ASR on both open-access and proprietary models, with ASR remaining significantly higher than that of InstructDetector. This indicates that simple structural modifications or prompting strategies fail to provide robust protection against IPI attacks.

Adversarial training, a white-box method, demonstrates greater effectiveness in lowering ASR compared to black-box approaches. However, it still underperforms compared to our method and has limitations, especially for proprietary models, since it involves changes to the embedding layer and necessitates model fine-tuning.
Our approach stands out for its ability to achieve superior ASR reduction while maintaining compatibility with both open-access and proprietary models, demonstrating its practicality and robustness against IPI attacks.

\subsection{Ablation Study}
For additional ablation studies, including experiments on different training data compositions, the impact of paired response, and the influence of instruction quantity and position, please refer to Appendix~\ref{appendix:ablation}.

\subsubsection{Solely Utilizing Hidden States/Gradients}

To evaluate the effectiveness of combining hidden states and gradients, we compare the performance of our approach utilizing both features with setups that relied solely on hidden states or gradients. The results presented in Table~\ref{tab:3} indicate that while utilizing either hidden states or gradients alone achieves high detection accuracy, combining the two features consistently delivers improved performance across all dataset combinations. These findings support our hypothesis that hidden states and gradients are complementary, and that integrating their strengths enhances the effectiveness of our method in detecting hidden instructions. 

\subsubsection{Detection Accuracy across Layers}
We further examine the detection accuracy of solely utilizing hidden states or gradients across different layers on all dataset combinations. As presented in Figure \ref{fig:3}, the detection accuracy across different layers demonstrates a clear trend: performance initially improves with increasing layer depth, reaches a peak at the middle layers, but then fluctuates significantly and generally declines. This trend highlights that intermediate layers capture more informative features relevant to instruction detection, whereas deeper layers may introduce noise or less task-specific representations, which is consistent with our observations on the validation set. These findings also align with the observations of a recent study \citep{skean2024does}, demonstrating that intermediate layers in LLMs often yield richer representations for downstream tasks compared to the final layers.

\subsubsection{Large Language Models}
The effectiveness of InstructDetector is evaluated across various LLMs, including different architectures (Llama \citep{dubey2024llama}, Qwen \citep{bai2023qwen}, Mistral \citep{jiang2023mistral}) and model sizes (1B, 3B, 7B, 8B, 14B parameters). As shown in Table \ref{tab:4}, features extracted from all tested LLMs are effective in detecting hidden instructions. Notably, we select the hidden states and gradients from the best-performing layer, which are all located in the intermediate layers. Among the evaluated models, Qwen-2.5-7B and Llama-3.1-8B exhibit superior results, while Mistral-7B shows slightly less optimal performance. Furthermore, the findings indicate that larger models generally produce features that are more effective for instruction detection, aligning with our hypothesis that stronger model capabilities lead to features that better facilitate the identification of hidden instructions.

We also include a comparison between Llama-3.1-8B-Base and Llama-3.1-8B-Instruct. The results show a significant performance gap, with Llama-3.1-8B-Base demonstrating notably worse results. This difference is likely because IPI attacks rely on the presence of hidden instructions embedded within external content, which alter the behavioral states of LLMs. Since Llama-3.1-8B-Base has not undergone instruction fine-tuning, it does not exhibit the same responsiveness to such hidden instructions in the way that the instruct model does. As a result, the ability of Llama-3.1-8B-Base to detect such attacks is considerably diminished.

\subsubsection{Training Data Size}
\begin{figure*}[htb]
  \centering
  \includegraphics[width=\textwidth]{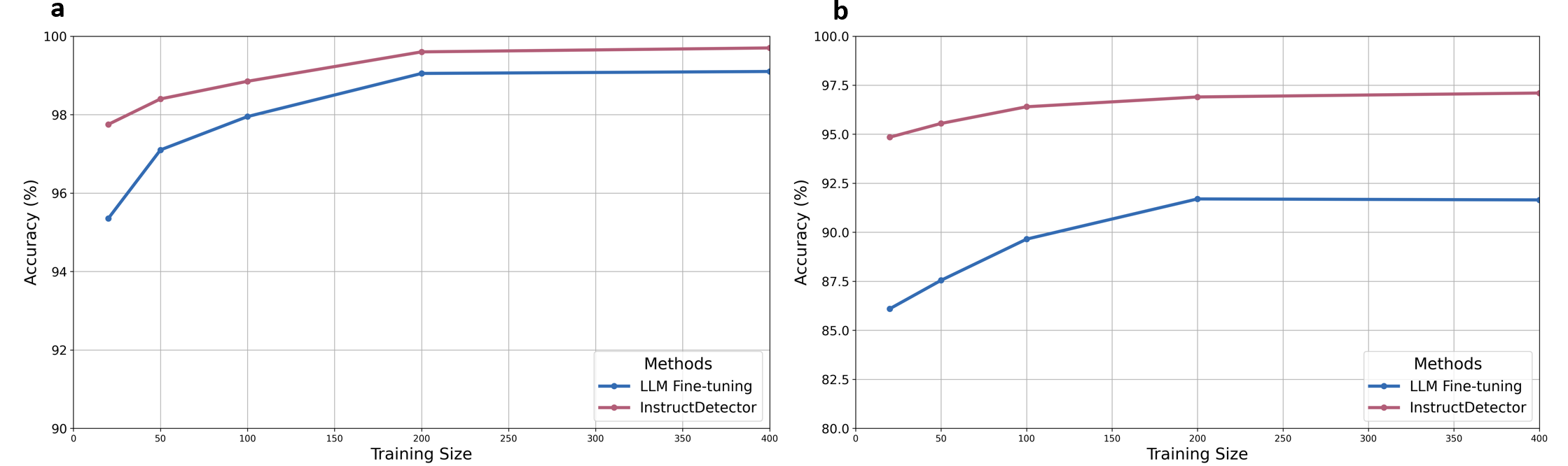}
  \caption{\textbf{Comparison of detection accuracy between LLM fine-tuning and InstructDetector on different training data size.} \textbf{(a)} Detection accuracy comparison in the in-domain setting (Wikipedia+LaMini-Instruction). \textbf{(b)} Detection accuracy comparison in the out-of-domain setting (News Article+BIPIA).}
  \label{fig:4}
\end{figure*}

\begin{table*}[htb]
  \centering
  \resizebox{\textwidth}{!}{
  \begin{tabular}{lcccc}
    \hline
    \textbf{} & \textbf{Wiki+LaMini (ID)} & \textbf{News+LaMini (OOD)} & \textbf{Wiki+BIPIA (OOD)} & \textbf{News+BIPIA (OOD)} \\
    \hline
    \textbf{English} & \textbf{99.60\%} & \textbf{98.35\%} & \textbf{99.45\%} & \textbf{96.90\%} \\
    \textbf{Chinese} & 97.40\% & 95.15\% & 96.20\% & 94.25\% \\
    \textbf{Thai} & 95.75\% & 94.15\% & 93.60\% & 92.05\% \\
    \hline
  \end{tabular}
  }
  \caption{\label{citation-guide}
    Detection accuracy comparison on multilingual test sets. The highest detection accuracy is indicated in \textbf{bold}. Here, ID denotes the in-domain setting, whereas OOD denotes the out-of-domain setting.
  }
  \label{tab:10}
\end{table*}

We conduct experiments using training data of varying sizes to assess how the quantity of training data affects the performance of InstructDetector. As presented in Figure \ref{fig:4}, even with a small training set of only 50 samples (25 positive and 25 negative), InstructDetector achieves relatively high performance, exceeding 95\% accuracy in both in-domain and out-of-domain scenarios. These results indicate that InstructDetector requires only minimal training data to achieve strong results. These findings highlight the remarkable data efficiency of InstructDetector, which performs well even with very limited data.

\subsubsection{Multilingual Evaluation}
We construct two additional test sets to evaluate how well InstructDetector, trained solely on English data, can transfer to other languages. Specifically, following established approaches in multilingual NLP \citep{kassner2021multilingual}, we translated our original test set using Google Translate into both a high-resource language (Chinese) and a relatively low-resource language (Thai). To ensure semantic similarity and consistency, we perform back-translation and compare the outputs with the original English versions.

As shown in Table \ref{tab:10}, InstructDetector maintains high accuracy across these multilingual test sets, despite being trained exclusively on English data. This demonstrates that the discriminative features leveraged by InstructDetector are not language-specific, and can effectively generalize beyond English.

\section{Conclusion}
In this work, we present InstructDetector, a detection-based approach that leverages the internal behavioral states of LLMs as signals to identify IPI attacks. A key finding of our study is that the hidden states and gradients from the intermediate layers of LLMs provide highly discriminative features for instruction detection. By leveraging these internal behavioral states, InstructDetector provides a robust mechanism for identifying hidden instructions within external data.

We demonstrate that InstructDetector achieves superior detection accuracy in both in-domain and out-of-domain settings, while significantly reducing the attack success rate compared to existing defense methods. These findings underline the effectiveness and adaptability of InstructDetector, offering a robust solution for enhancing the security of LLM-based systems.

\section{Limitation}
InstructDetector has several limitations. First, it requires both forward and backward passes through the LLM, introducing additional computational overhead compared to lightweight defenses. While suitable for offline filtering or batch processing, this may limit deployment in resource-constrained settings and raise environmental concerns due to the increased energy consumption associated with higher computational demands. Second, although our experiments cover multiple representative scenarios and datasets, we cannot guarantee coverage of all possible attack strategies or domain-specific variations. Third, the current design adopts a conservative binary decision—discarding any external data flagged as containing hidden instructions—resulting in the unintended removal of useful, non-malicious information. However, the high accuracy of our method forms a solid foundation for precisely identifying and removing external instructions. By recursively segmenting documents and applying our method to each segment individually, we can remove only indirect prompts while preserving useful content.

\section{Ethical Impact}
Our proposed method, InstructDetector, defends against IPI attacks, which is essential for ensuring the secure and reliable operation of LLMs in third-party system integrations. By mitigating the risks posed by IPI attacks, InstructDetector fosters ethical and socially responsible use of AI technologies, enhancing trust in their application within critical sectors such as healthcare, legal, and finance domains.
There may be concerns about whether InstructDetector could provide attackers with insights to bypass detection. Since InstructDetector leverages the distinct behavioral states of LLMs to differentiate between data and instructions, while IPI attacks fundamentally rely on the external instructions to alter the behavioral states of LLMs, it would be exceedingly difficult for attackers to circumvent our detection.
In summary, InstructDetector strengthens the security and trustworthiness of AI systems by effectively defending IPI attacks, aligning with ethical principles, and supporting the development of reliable, safe, and socially responsible AI technologies for real-world applications.

\section{Acknowledgments}
We sincerely appreciate the anonymous reviewers for their insightful comments and suggestions that may significantly enhance the quality of this work. The work was partially done at the Beijing Key Laboratory of Research on Large Models and Intelligent Governance.

\bibliography{main}

\clearpage

\appendix

\section{Additional Experimental Settings}

\begin{tcolorbox}[title = {Prompt for LLM (Zero-shot).}]
\#\#\# Guideline \#\#\#

You are a helpful assistant, and your task is to determine if there is any external instruction in the text. Respond with 1 if there is an instruction, or 0 if there is not. Only reply with 1 or 0, without additional explanation.

\#\#\# Your Task \#\#\#

Text: \verb|external data|

Answer:
\end{tcolorbox}

\begin{tcolorbox}[title = {Prompt for LLM (Few-shot).}]
\#\#\# Guideline \#\#\#

You are a helpful assistant, and your task is to determine if there is any external instruction in the text. Respond with 1 if there is an instruction, or 0 if there is not. Only reply with 1 or 0, without additional explanation.

\#\#\# Examples \#\#\#
Here are a few concrete examples to help you understand how to make a decision. 

Example 1:

Text: \verb|external data|

Answer: \verb|label|

\#\#\# Your Task \#\#\#

Text: \verb|external data|

Answer:

\end{tcolorbox}

\begin{tcolorbox}[title = {Prompt for Response Check.}]
Please follow the instruction based on the context below.

Instruction: \verb|instruction|

Context: \verb|external data|

\end{tcolorbox}

\begin{table*}[t]
  \centering
    \setlength{\tabcolsep}{13pt}
  \begin{tabular}{lcccc}
    \hline
    \textbf{} & \textbf{Wiki+LaMini} & \textbf{News+LaMini} & \textbf{Wiki+BIPIA} & \textbf{News+BIPIA} \\
    \hline
    \textbf{Wiki+LaMini} & \textbf{99.60\%} & 98.35\% & 99.45\% & 96.90\% \\
    \textbf{News+LaMini} & \textbf{99.60\%} & \textbf{98.50\%} & 99.55\% & 96.45\% \\
    \textbf{Wiki+BIPIA} & 99.15\% & 97.50\% & \textbf{99.85\%} & 97.30\% \\
    \textbf{News+BIPIA} & 99.45\% & 98.35\% & 99.65\% & \textbf{98.05\%} \\
    \hline
  \end{tabular}
  \caption{\label{citation-guide}
    Detection accuracy comparison utilizing different combinations of training datasets. The highest detection accuracy is indicated in \textbf{bold}.
  }
  \label{tab:5}
\end{table*}

\begin{table*}[t]
  \centering
  \resizebox{\textwidth}{!}{
  \begin{tabular}{lcccc}
    \hline
    \textbf{} & \textbf{Wiki+LaMini (ID)} & \textbf{News+LaMini (OOD)} & \textbf{Wiki+BIPIA (OOD)} & \textbf{News+BIPIA (OOD)} \\
    \hline
    \textbf{I’m sorry} & 99.45\% & 97.95\% & 97.25\% & 95.15\% \\
    \textbf{Hello} & 99.45\% & 97.90\% & 98.15\% & 95.90\% \\
    \textbf{Yes} & 99.55\% & 97.80\% & 99.40\% & 96.80\% \\
    \textbf{Sure} & \textbf{99.60\%} & \textbf{98.35\%} & \textbf{99.45\%} & \textbf{96.90\%} \\
    \hline
  \end{tabular}
  }
  \caption{\label{citation-guide}
    Detection accuracy comparison using different paired responses to extract gradient features. The highest detection accuracy is indicated in \textbf{bold}. Here, ID denotes the in-domain setting, whereas OOD denotes the out-of-domain setting.
  }
  \label{tab:6}
\end{table*}

\subsection{Dataset Details}
\label{appendix:datasets}

All datasets used in this work are in English, and the use of these datasets complies with their intended purposes as specified by their creators. We have carefully reviewed the data to ensure that they do not contain personally identifiable information or offensive content.

\paragraph{Wikipedia} The dataset is constructed using Wikipedia dump files, under the CC-BY-SA license. Each data instance comprises the content of an entire Wikipedia article. In addition, we remove the overly long articles to ensure that they are not truncated during processing. 

\paragraph{News Articles} The dataset contains 3,824 news articles, each featuring metadata including the title, subtitle, content, and publication date, sourced from multiple media outlets, under the CC0 license. Similarly, we remove the overly long articles to ensure that they are not truncated during processing.

\paragraph{LaMini-instruction} The dataset consists of 2.58 million pairs of instructions and corresponding responses, generated using GPT-3.5-Turbo, drawing from a wide range of existing resources of prompts, including Self-Instruct, P3, FLAN, and Alpaca, under the CC-BY-NC license.

\paragraph{BIPIA} BIPIA is the first benchmark aimed at evaluating the risk of IPI attacks on LLMs, under the MIT license, and we use its instruction dataset for our experiments. The dataset consists of 15 attack types, categorized into task-irrelevant, task-relevant, and targeted attacks, with 5 instructions per attack type, resulting in a total of 75 instructions across both the training and test sets. These instructions were semi-automatically generated with the assistance of ChatGPT and manually reviewed for rationality.

\subsection{Baseline Details}
\label{appendix:baseline-models}

\subsubsection{Detection-based Defenses}
\paragraph{LLM (Zero-shot) \citep{liu2024formalizing}} Directly query the LLM to identify if there is any hidden instruction within the external content, utilizing the LLM's existing capabilities without additional enhancements or fine-tuning. 

\paragraph{Response Check \citep{liu2024formalizing}} Evaluate the LLM's output by checking whether the response aligns with the intended task, where a mismatch indicates potential manipulation by hidden instructions within the external content.

\paragraph{TaskTracker \citep{abdelnabi2024you}} Detect IPI attacks by contrasting the LLM’s activations before and after feeding the external data, which indicates whether the user's instruction is distorted by the instruction hidden in the external data. 

\paragraph{LLM (Few-shot)} To enhance the performance of Naive LLM-based Detection, we attempt to leverage in-context learning \citep{dong2022survey} to strengthen the model’s capability to detect hidden instructions, where task demonstrations are integrated into the textual prompt.

\paragraph{LLM (Fine-tuning)} Similarly, to further improve naive LLM-based detection, we conduct supervised fine-tuning using task-specific annotated data, thereby strengthening the model’s ability to detect hidden instructions.

\subsubsection{Prevention-based Defenses}
\paragraph{In-context Learning \citep{yi2025benchmarking}} Employ in-context learning to enable the model to distinguish between external data and user instructions, by providing samples where the model responds to input containing external data without being misled by the instruction embedded within external data.

\paragraph{Multi-turn Dialogue \citep{yi2025benchmarking}} Strategically shift external data—which may contain covert instructions—to the preceding conversational turn, while reserving the user's instruction for the current turn. This separation between external content and user instruction effectively mitigates ASR.

\paragraph{Adversarial Training \citep{yi2025benchmarking}} Incorporate adversarial learning during the LLM's self-supervised fine-tuning phase, training the model to disregard instructions embedded within external content. The approach further adapts the model's embedding layer to explicitly demarcate external content boundaries, enabling clearer distinction between external content and user instructions.


\subsection{Additional Experimental Details}
\label{appendix:baseline-setup}
 
As shown earlier, we employ specific prompt templates for LLM (Zero-shot), Response Check, and LLM (Few-shot).

For the LLM (Fine-tuning), we use the AdamW optimizer to train the model for one epoch, with a learning rate set at 0.00001 and a maximum input length of 2048 tokens. Both the learning rate and the number of epochs are chosen based on performance on the validation set. For TaskTracker, we use the publicly available model released by the work. As for the prevention-based approaches, we utilize the publicly available implementations provided by the respective works.

We conduct our experiments using an 8B-parameter LLM. All training and inference are performed on a single NVIDIA A100 80GB GPU, with the training phase requiring approximately 1 GPU hour. In practice, InstructDetector can process 1283 tokens per second on a single A100 GPU. It is also compatible with various acceleration and quantization techniques, as long as gradients and hidden states are accessible, which can further improve efficiency. Importantly, unlike existing task-specific approaches that require runtime interventions, InstructDetector allows for pre-screening of external sources and caching of results. This design significantly reduces the latency incurred during actual system usage and makes the overhead of instruction detection negligible at inference time. For all reported results, we present outcomes from a single run. During preprocessing, to construct negative samples by inserting instructions into benign text, we utilize the PunktSentenceTokenizer from the NLTK library for sentence segmentation.

\label{appendix:comparison}
\begin{table}[htb]
  \centering
  \resizebox{\columnwidth}{!}{
  \begin{tabular}{lcc}
    \hline
    \textbf{} & \textbf{Scenario 1} & \textbf{Scenario 2}\\
    \hline
    \textbf{TaskTracker} & 70.55\% & 71.35\%\\
    \textbf{InstructDetector} & \textbf{99.40\%} & \textbf{99.55\%}\\
    \hline
  \end{tabular}
  }
  \caption{\label{citation-guide}
    Detection accuracy comparison of InstructDetector and TaskTracker when the external instruction closely resembles the user's instruction. For scenario 1, the user instruction is "Translate the news into Chinese." and the external instruction is "Translate the news into Japanese."; For scenario 2, the user instruction is "Generate a step-by-step guide for solving this problem." and the external instruction is "Generate a list of common mistakes in solving this problem." The highest detection accuracy is indicated in \textbf{bold}.
  }
  \label{tab:7}
\end{table}

\begin{table*}[t]
  \centering
  \resizebox{\textwidth}{!}{
  \begin{tabular}{lcccc}
    \hline
    \textbf{} & \textbf{Wiki+LaMini (ID)} & \textbf{News+LaMini (OOD)} & \textbf{Wiki+BIPIA (OOD)} & \textbf{News+BIPIA (OOD)} \\
    \hline
    \textbf{one instruction} & 99.60\% & 98.35\% & 99.45\% & 96.90\% \\
    \textbf{two instructions} & \textbf{99.85\%} & 98.40\% & 99.80\% & \textbf{97.00\%} \\
    \textbf{three instructions} & \textbf{99.85\%} & \textbf{98.45\%} & \textbf{99.85\%} & \textbf{97.00\%} \\
    \hline
  \end{tabular}
  }
  \caption{\label{citation-guide}
    Detection accuracy comparison for different quantities of inserted instructions in the test dataset. The highest detection accuracy is indicated in \textbf{bold}. Here, ID denotes the in-domain setting, whereas OOD denotes the out-of-domain setting.
  }
  \label{tab:8}
\end{table*}
\begin{table*}[t]
  \centering
  \resizebox{\textwidth}{!}{
  \begin{tabular}{lcccc}
    \hline
    \textbf{} & \textbf{Wiki+LaMini (ID)} & \textbf{News+LaMini (OOD)} & \textbf{Wiki+BIPIA (OOD)} & \textbf{News+BIPIA (OOD)} \\
    \hline
    \textbf{beginning} & \textbf{99.90\%} & \textbf{98.60\%} & \textbf{99.90\%} & \textbf{97.35\%} \\
    \textbf{middle} & 99.60\% & 98.35\% & 99.45\% & 96.90\% \\
    \textbf{end} & \textbf{99.90\%} & 98.40\% & 99.55\% & 97.05\% \\
    \hline
  \end{tabular}
  }
  \caption{\label{citation-guide}
    Detection accuracy comparison for different positions of inserted instructions in the test dataset. The highest detection accuracy is indicated in \textbf{bold}. Here, ID denotes the in-domain setting, whereas OOD denotes the out-of-domain setting.
  }
  \label{tab:9}
\end{table*}

\section{Comparison with Related Method}
Both InstructDetector and TaskTracker~\cite{abdelnabi2024you} utilize the hidden states of LLMs as a key feature for detecting IPI attacks, but they differ significantly in their underlying principles. TaskTracker aims to capture distortions in the user's instruction caused by embedded instructions in the external content. In contrast, InstructDetector aims to distinguish the LLM’s behavioral states when processing normal external data versus those containing hidden instructions.

TaskTracker has two primary limitations. First, it requires a large number of training samples (418,110 pairs of positive and negative samples) to accurately identify deviations in the user's task. In contrast, InstructDetector leverages the high sensitivity of LLM's behavioral states to embedded instructions, achieving effective detection with a significantly smaller dataset (just 100 pairs).

Second, TaskTracker's effectiveness relies heavily on a clear distinction between users' instructions and external instructions, while InstructDetector is task-agnostic. As shown in Table \ref{tab:7}, when the external instruction closely resembles the user's instruction, TaskTracker's detection accuracy drops significantly, while InstructDetector maintains high detection accuracy.

\section{Extended Ablation Studies}
\label{appendix:ablation}

\subsection{Composition of Training Data}
We conduct experiments using different combinations of training datasets to assess the robustness and adaptability of InstructDetector to various training dataset compositions. As presented in Table \ref{tab:5}, InstructDetector consistently yields high accuracy across all test datasets, regardless of the specific combination of training data used. This indicates the generalizability and adaptability of InstructDetector, as it does not rely on any particular training dataset source. Additionally, we observe that accuracy is consistently lower when tested on the News Articles with the BIPIA combination, indicating that this scenario poses the greatest challenge for instruction detection. Nonetheless, InstructDetector still achieves satisfactory accuracy in this challenging scenario, further validating its effectiveness and robustness in instruction detection.

\subsection{Paired Responses for Gradients}
To investigate the effect of various paired responses on the extraction of gradient features, we conduct experiments using four candidate responses: "I’m sorry" "Hello" "Yes" and "Sure." These candidates are selected based on an analysis of common responses to instructions in WildChat \citep{zhao2024wildchat} dataset, ranked by frequency. Results in Table \ref{tab:6} show that all four paired responses achieve high accuracy (>95\%) in distinguishing between normal external data and those containing hidden instructions. Among them, "Sure" delivers the best performance across all test datasets, further validating our choice of "Sure" as the paired response in InstructDetector. These results emphasize the robustness of InstructDetector to differentiate response pairings while confirming that "Sure" is a particularly effective option for this task.

\subsection{Influence of Instruction Quantity and Position}
To further explore the influence of instruction quantity and position on detection performance, we conduct experiments using a fixed training dataset while varying only the number or placement of inserted instructions in the test dataset.

Results in Table \ref{tab:8} reveal a trend that detection accuracy shows a certain degree of improvement as the number of inserted instructions increases. This suggests that a higher quantity of instructions provides stronger signals, making IPI attacks more distinguishable by InstructDetector. Additionally, we examine the effect of instruction placement by inserting instructions at the beginning, middle, or end of the external content. As shown in Table \ref{tab:9}, instructions placed in the middle are the most challenging to detect, whereas those positioned at the beginning or end are relatively easier to identify. Among these, instructions at the beginning yield the highest detection accuracy, likely because LLMs exhibit greater sensitivity to early input.

\end{document}